\documentclass[preprint,prb,aps,showpacs]{revtex4}
\usepackage{amsmath,amssymb}
\usepackage{hyperref}
\usepackage{graphics,psfrag}
\usepackage{graphicx,psfrag}
\usepackage{epsfig}
\usepackage{subfig}
\usepackage{float}

\expandafter\ifx\csname package@font\endcsname\relax\else
 \expandafter\expandafter
 \expandafter\usepackage
 \expandafter\expandafter
 \expandafter{\csname package@font\endcsname}
\fi

\begin{document}

\title{Emergence of competing magnetic interactions induced by Ge doping in the semiconductor FeGa3}

\author{J. C. Alvarez-Quiceno}
\author{M. Cabrera-Baez}
\author{R. A. Ribeiro}
\author{M. A. Avila}
\email{avila@ufabc.edu.br}
\author{G. M. Dalpian}
\email{gustavo.dalpian@ufabc.edu.br}
\affiliation{Centro de Ci\^encias Naturais e Humanas, Universidade Federal do ABC, Santo Andr\'e, SP, Brazil}

\author{J. M. Osorio-Guill\'en}
\email{mario.osorio@udea.edu.co}
\affiliation{Instituto de F\'{\i}sica, Universidad de Antioquia UdeA, Calle 70 No 52-21, Medell\'{\i}n, Colombia}
\affiliation{Centro de Ci\^encias Naturais e Humanas, Universidade Federal do ABC, Santo Andr\'e, SP, Brazil}

\begin{abstract}
FeGa$_3$ is an unusual intermetallic semiconductor that presents intriguing magnetic responses to the tuning of its electronic properties. When doped with Ge, the system 
evolves from diamagnetic to paramagnetic to ferromagnetic ground states that are not well understood. In this work, we have performed a joint theoretical and experimental 
study of FeGa$_{3-x}$Ge$_x$ using Density Functional Theory and magnetic susceptibility measurements.
For low Ge concentrations we observe the formation of localized moments on some Fe atoms and, as the dopant concentration increases, a more delocalized magnetic 
behavior emerges.
The magnetic configuration strongly depends on the dopant distribution, leading even to the appearance of antiferromagnetic interactions in certain configurations.
\end{abstract}


\maketitle

\section{Introduction}

The physics of Fe-based intermetallic semiconductors such as FeSi, FeSb$_2$ and FeGa$_3$ can present very unusual phenomena, creating an important framework for 
the development of condensed matter physics. 
These materials can present properties of strongly correlated electron systems, owing to the presence of the Fe $d$ levels, concurrent with 
semiconducting gaps due to the strong hybridization between the Fe $d$ and the post-transition metal or metalloid $s,p$ levels.
They have been studied mainly for their thermoelectric properties\cite{Amagai2004,Haldolaarachchige2011,Wagner-Reetz2014}.
However, they are also interesting from fundamental physical aspects\cite{Osorio-Guillen2012}, still presenting important unsolved puzzles.

Here we  address  the puzzling magnetic response upon electronic doping of FeGa$_3$, one of the few Fe-based materials known to be diamagnetic.
This diamagnetism of the pure compound has also been assigned to the strong hybridization between the Fe $3d$ levels and the Ga $4s$ and $4p$ levels.
Starting from this already unusual state, strikingly different magnetic responses can be observed upon chemical substitutions that lead to doping with holes or electrons\cite{Cabrera-Baez2013}. 

There is much controversy in the literature regarding the most stable magnetic configurations of pure and doped FeGa$_3$, as well as on the mechanisms involved.
Some works propose an antiferromagnetic ground state for pure and doped FeGa$_3$ \cite{Pickett2010,Gamza2014},
whereas others report a diamagnetic ground state\cite{Osorio-Guillen2012,Tsujii2008,Hadano2009}.
There are also authors invoking an itinerant origin for the ferromagnetic ground state \cite{Singh2013}, whereas other reports show
a localized nature for the magnetic moments\cite{Bittar2010,Kotur2013}, or both \cite{Gippius2014}.

In order to understand this system, we have performed a joint theoretical and experimental work using First-principles Density Functional Theory (DFT)
calculations and bulk magnetization techniques on Ge-doped FeGa$_3$ single crystals. 
Our results confirm that, upon doping, the system quickly evolves to a mainly ferromagnetic (FM) ground state.
However, the evolution of magnetic response to Ge doping is far from trivial. 
The magnetic moments are not evenly distributed throughout all Fe atoms as expected in an itinerant picture, and different groupings of magnetic moments on the Fe atoms 
are resolved, that 
depend on both doping concentration and dopant distribution. 
Additionally, some distributions do indeed lead to Fe moments aligned antiparallel to the mostly FM state. 
Our results point to an evolution from localized to delocalized character and an intricate competition between FM \emph{vs.} antiferromagnetic (AFM) pictures for the 
magnetic response in the doped 
materials, which may lead to the conclusion that almost all of the previous works were providing partially correct but incomplete descriptions of this fascinating system.

\section{Methodology}

\subsection*{Crystal growth and experimental characterizations}

Polyhedral single crystals of FeGa$_{3-x}$Ge$_x$ with $x=0.27$ were grown by a standard Ga self-flux method~\cite{Umeo2012, Ribeiro2012}. 
High purity elements were sealed inside evacuated silica ampoules with a Fe:Ga:Ge proportion of 1:[3(3-x)]:[3x]. 
These ampoules were heated to 1100~$^\circ$C and cooled to 550~$^\circ$C over 150~h, then removed from the furnace and quickly centrifuged for separation of the 
molten flux. 
The effective doping level of each crystal was estimated by comparing the Sommerfeld coefficient, transition temperature and effective moment with the results by Umeo \emph{et al.}~\cite{Umeo2012}. 
X-ray diffraction patterns of powdered crystals were obtained on a Bruker D8 Focus machine and are all consistent with the FeGa$_{3}$ type structure, with refined lattice 
parameters of the pure 
FeGa$_{3}$ in agreement with published works~\cite{Umeo2012,Bittar2010}. 
DC magnetic susceptibility for temperatures between 2 and 300~K was measured on a Quantum Design SQUID-VSM. 
For applied fields below 30~Oe the flux gate accessory was used to cancel any remnant field in the superconducting magnet, then a sequence of zero field cooled (ZFC), 
field-cooled cooling (FCC) and field cooled warming (FCW) branches was performed.

\subsection*{Density Functional Theory calculation}

Spin-polarized first-principles calculations based on DFT were carried out using the Generalized Gradient Approximation
with a parametrization targeted especially for solids, namely PBEsol~\cite{Perdew2008}. Also, we performed calculations introducing an
on-site Coulomb interaction term, the DFT+U method~\cite{Dudarev:1998p2091,Bengone:2000p316}, characterized by the Hubbard $U$ parameter acting on Fe $d$-states.  
The Kohn-Sham equations were solved using the Projected Augmented Wave (PAW) method~\cite{Blochl1994}, as implemented in the
Vienna {\it ab-initio} Simulation Package (VASP)~\cite{Kresse1996,Kresse1999}. The PAW atomic reference configurations are
$3p^63d^64s^2$, $3d^{10}4s^24p$, $3d^{10}4s^24p^2$ for the Fe, Ga and Ge respectively. These settings reproduce correctly the 
properties of this material as reported in Ref.~\onlinecite{Osorio-Guillen2012}. The kinetic energy cut-off for the plane-wave expansion was 400 eV. 
For each dopant configuration, the lattice parameters were fixed at the FeGa$_3$ equilibrium crystal structure while the atomic positions were allowed to relax
without any symmetry constraint.
In the relaxation process, Monkhorst-Pack $\mathbf{k}$-point meshes were used, and depending of the size of the
supercell we used $\Gamma$-centered $\mathbf{k}$-point meshes of
$6\times6\times6$, $6\times6\times3$, and $2\times 2\times2$ for the calculation of charge and magnetization densities.

\section{Results and Discussion} 

FeGa$_3$ has a tetragonal crystal structure belonging to the space group $P4_{2}/mnm$ (136) with one Fe atom
located at the Wyckoff position $4f (u,u,0)$ and two types of Ga atoms: Ga1 occupies the Wyckoff position $4c(0, \frac{1}{2}, 0)$ and
Ga2 is located at the Wyckoff position $8j(u,u,w)$ \cite{Haussermann2002}.

\begin{figure}
\begin{center}
  \subfloat[$0.06e/$f.u.]{\includegraphics[width=5.4cm, height=5.4cm]{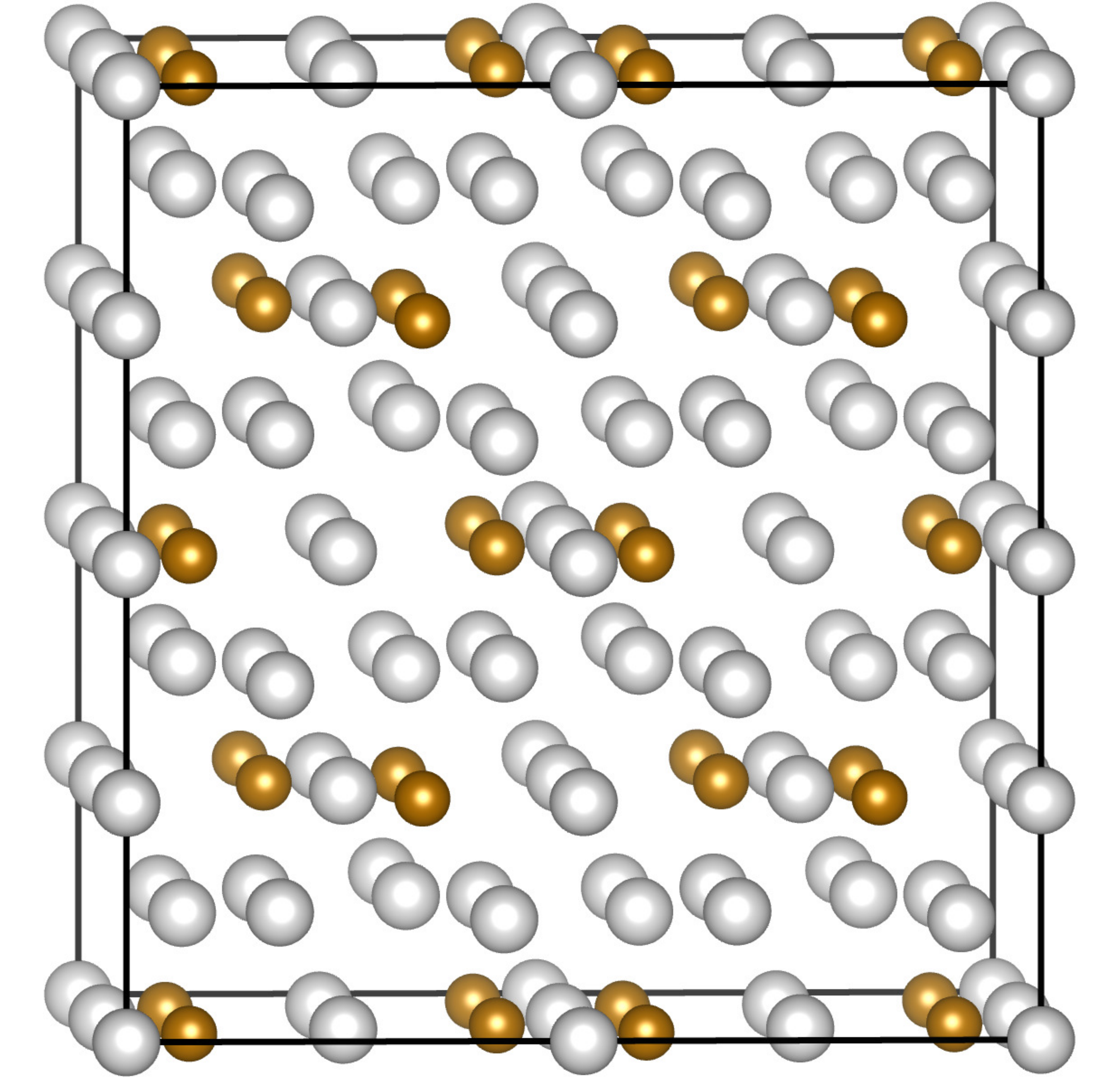}}
  \subfloat[$0.13e/$f.u.]{\includegraphics[width=5.4cm, height=5.4cm]{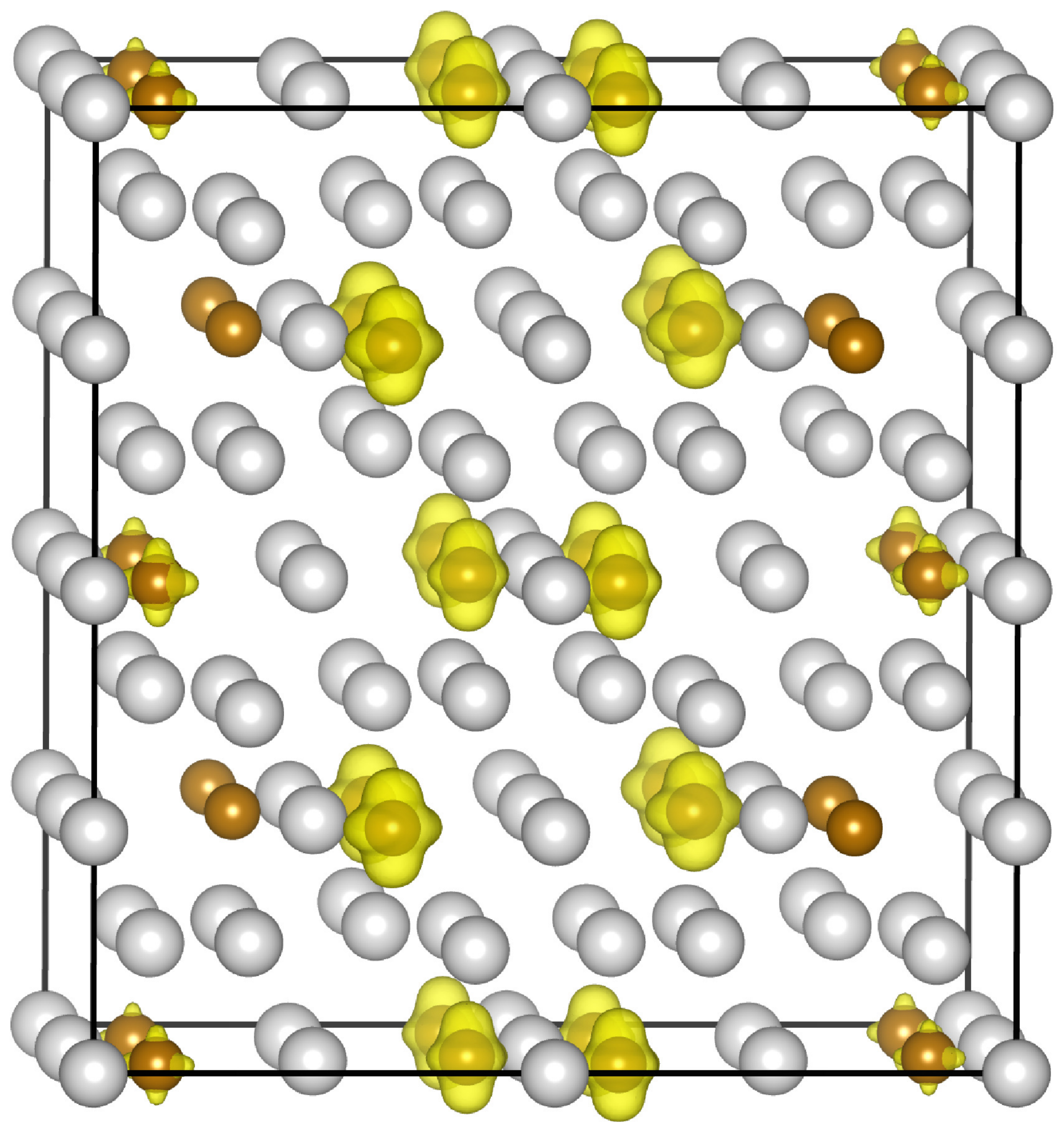}}
  \subfloat[$0.25e/$f.u.]{\includegraphics[width=5.4cm, height=5.4cm]{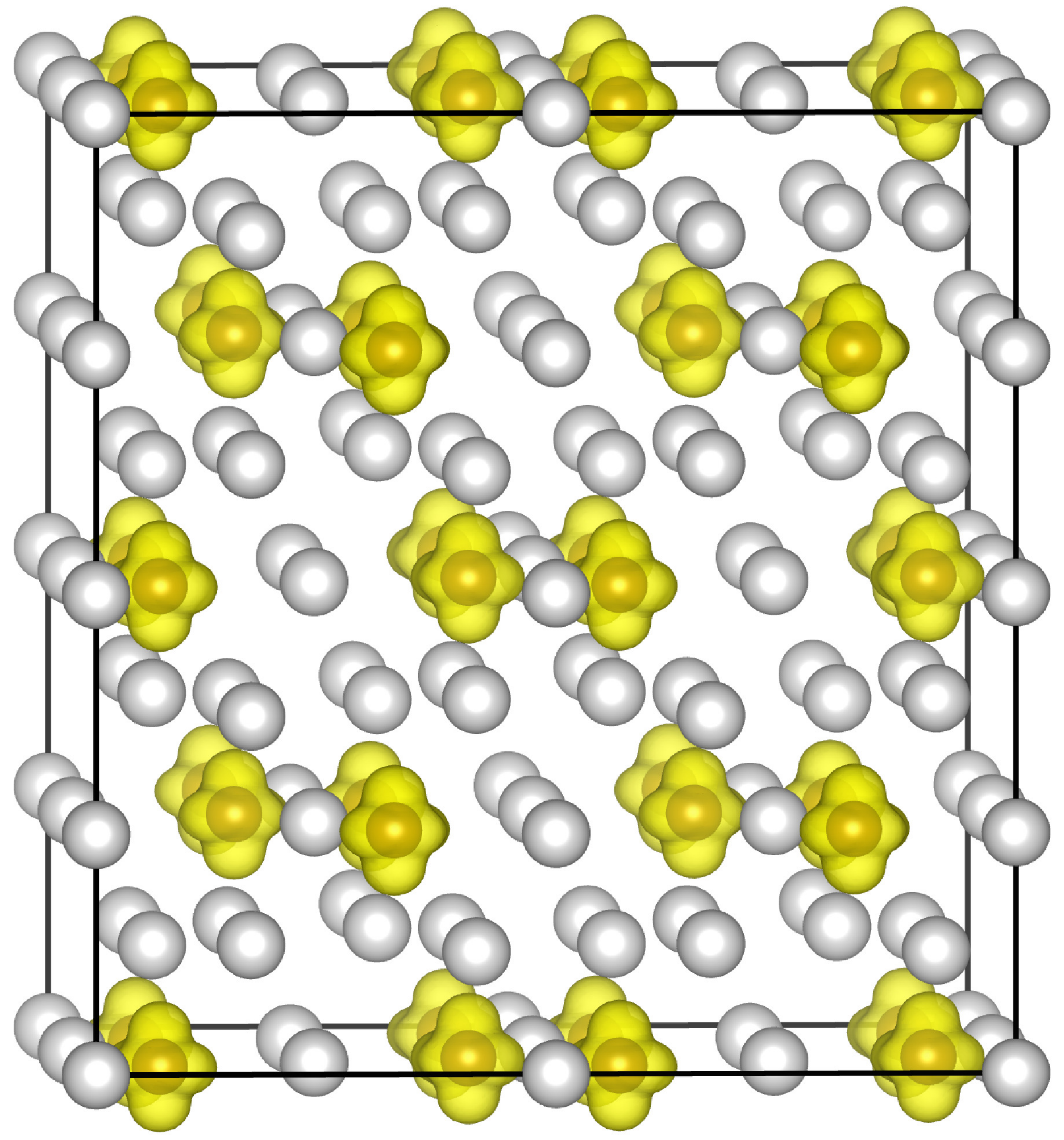}}
\end{center}
\caption{Electron doped FeGa$_3$ magnetization density of a $2\times2\times2$--supercell:
(a) small electron density (0.06e/f.u.), (b) medium electron density (0.13e/f.u.), and (c) large electron density (0.25e/f.u.).
Brown and grey spheres represents Fe and Ga atoms, respectively. The yellow surface represents the magnetization density around each atom.}
\label{Figure1}					
\end{figure}

The simplest way to theoretically model $n$-type doping in FeGa$_3$ is by shifting the Fermi level of the system up,
which will represent the occurrence of electrons in the conduction band. This preliminary approach leads to the observation that the magnetic response depends on the 
electron concentration on the system. For low electron densities ($\leq 0.09 e/$f.u.) we do not observe any perturbation of the system, and
no magnetic moments are observed in the Fe atoms. As we increase the electron density, 
we start to observe a magnetic moment in some Fe atoms, i.e., the electrons are not homogeneously
distributed through the whole lattice. In this case, the electrons seem to be self-trapped in a lattice distortion.
A similar self-trapping behavior has been reported before for the semiconductor MnO~\cite{Pang2012}.
At higher electron concentration ($0.25 e/$f.u.) it is observed a more delocalized (itinerant) 
magnetic state with similar magnetic moments in all Fe atoms (see Fig.~\ref{Figure1}). Test calculations have shown that these results 
are robust with respect supercell sizes.

After modeling the system with a preliminary effective doping model, we advance our calculations to an explicit doping
approach of FeGa$_3$ with Ge atoms (FeGa$_{3-x}$Ge$_x$). For this study, the host system is seen as a layered structure with the Ga1 ions placed at the same plane of 
Fe and Ga2 occupying its own separate plane.
Calculations were performed at concentrations $x$ = 0.03, 0.06, 0.09, 0.13, 0.19, 0.25, 0.38 and 0.50 for several
different distributions of substitutional Ge impurities throughout the supercell, always on Ga sites, with the appropriate amounts
to obtain the desired concentrations. We have also converged different spin configurations for each atomic distribution, including
(i) non-spin polarized, (ii) all Fe atoms with parallel spin and (iii) all Fe dimers with antiparallel spins. 
Along this work we will discuss the most stable spin configuration at each case.

When a single Ge atom is inserted in a $2\times2\times2$ supercell at either Ga1 or Ga2 site ($x = 0.03$), the Ga2 site is preferred by an energy difference of 0.11~eV.
In this case, the Ge atom also induces a tiny magnetic moment on its Fe neighbors ($5 \times 10^{-3}\mu_B$/f.u.), whereas
at the Ga1 site it does not lead to any magnetic moment on its Fe neighbors.
This contrast demonstrates the complex nature of this material, since the exact lattice position where each Ge atom enters directly influences the magnetic configuration, and 
actual samples will inevitably feature a statistical distribution of the dopant among the nearest neighbors of an Fe atom.
Consequently, a complete description of the magnetic behavior can only be achieved through a meticulous and thorough exploration
involving many different distributions, which we undertake next.

\begin{figure}
\begin{center}
\subfloat[PBEsol]{\includegraphics[width=0.68\textwidth]{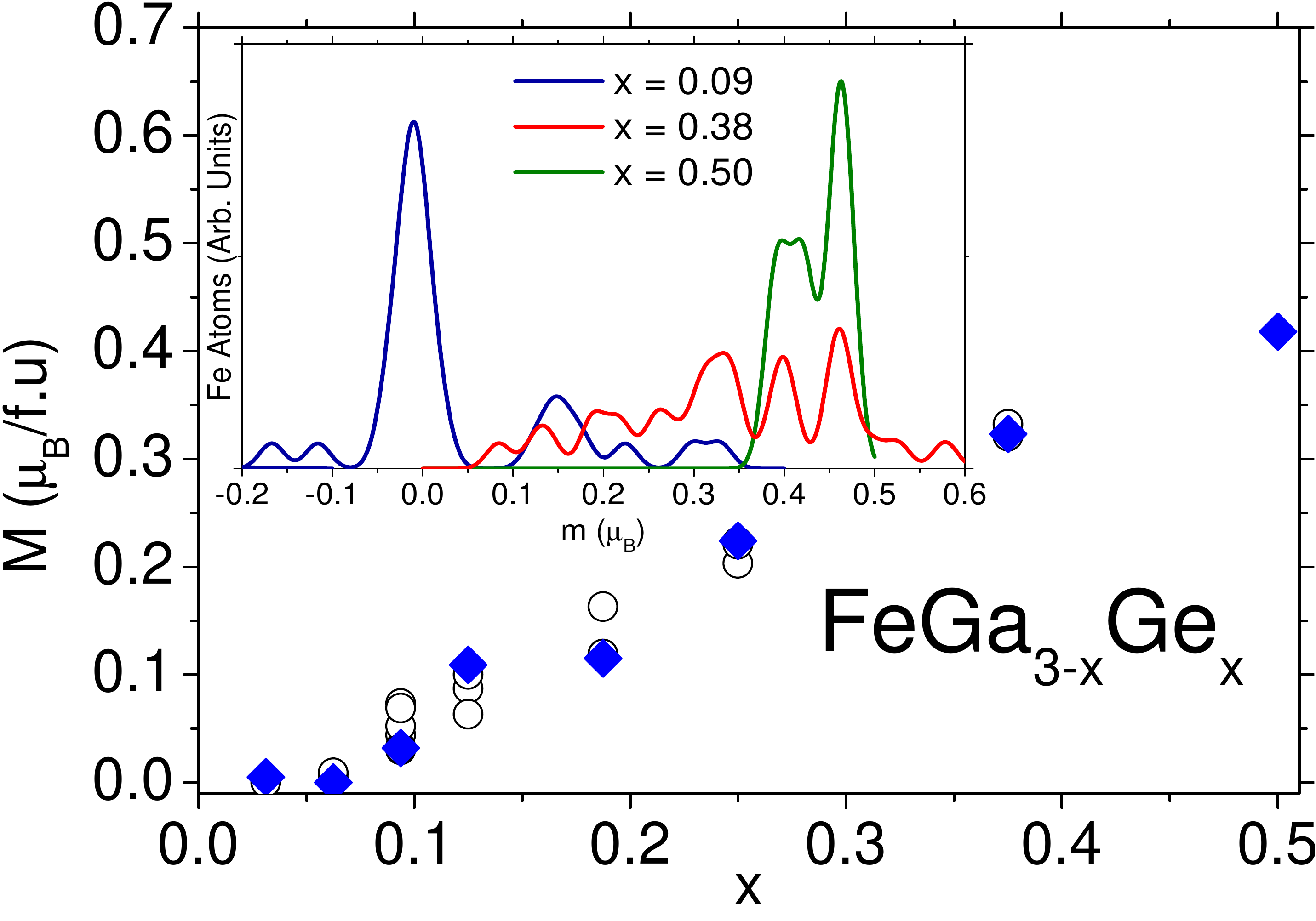}}  \\
\subfloat[DFT+U]{\includegraphics[width=0.65\textwidth]{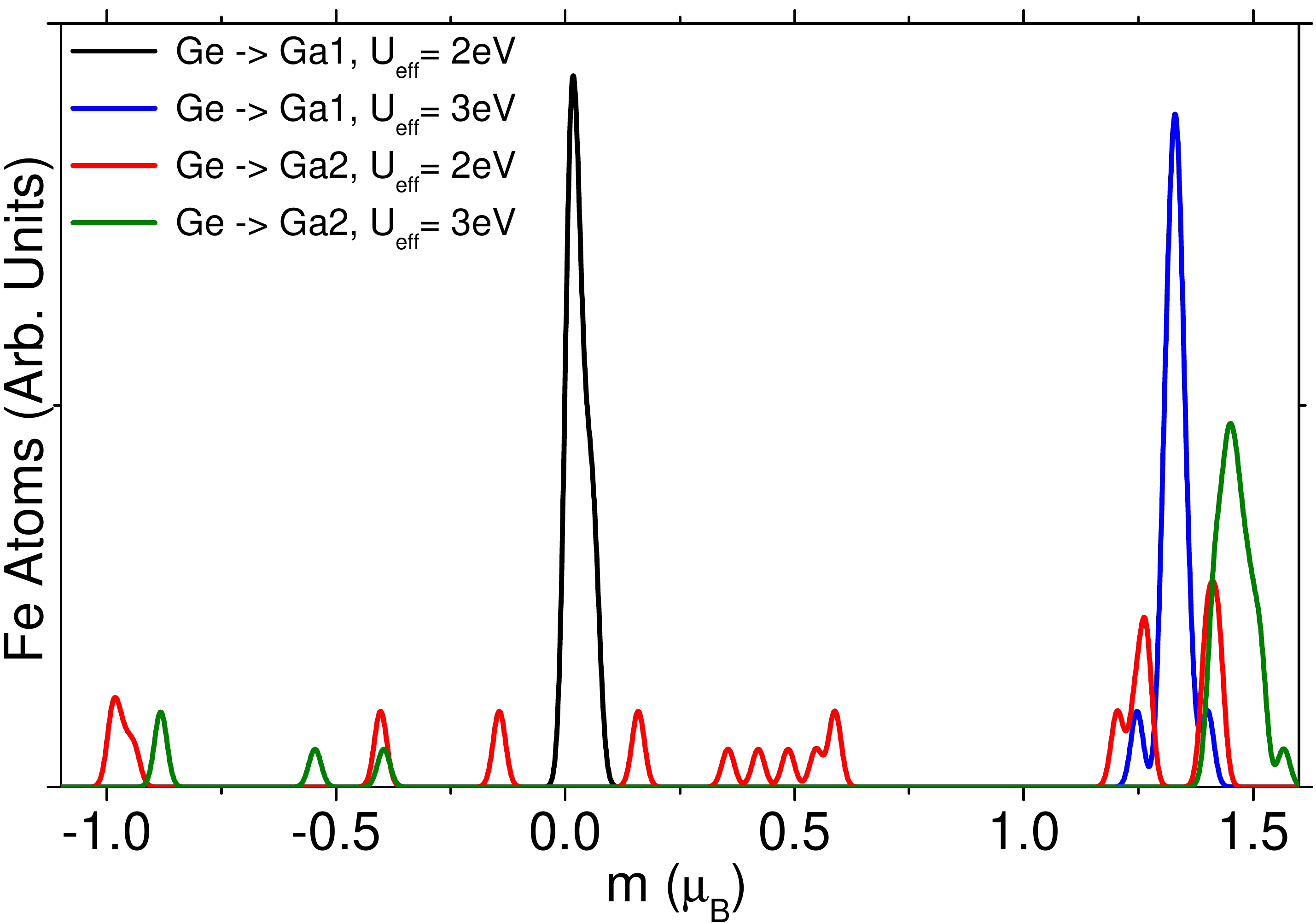}}
\end{center}
\caption{(a) PBEsol FeGa$_{3-x}$Ge$_x$ average magnetization per f.u. as a function of the Ge concentration $x$ for different substitutional distributions.
The blue diamonds indicate the lowest energy configuration for each concentration.
Inset: Fe magnetic moment distribution of the lowest energy configurations for $x =$ 0.09, 0.38 and 0.50.
(b) DFT+U Fe magnetic moment distribution for a $2\times2\times2$--supercell where one substitutional impurity is located
in either Ga1 or Ga2 site ($x=0.03$, U$_{\mathrm{eff}}$ = 2 and 3 eV).}
\label{Figure2}
\end{figure}

By increasing the dopant concentration in different supercells, we were able to obtain several different magnetic configurations,
wherein the total moment per f.u. also resulted different.
For a given concentration, there can be different spin configurations that differ simply by the lattice distribution of the Ge atoms.
This is exemplified in the main panel of Fig. \ref{Figure2}(a), where the total magnetic moment per f.u. is plotted
for several different distributions of a fixed dopant concentration. The blue diamonds in this figure indicate the most stable spin configuration.
For $x$ = 0.03 and 0.06 there exist distributions with null magnetic moment. For each concentration and distribution, we also obtain different values of 
magnetic moment on the Fe atoms, depending on their positions with respect to the dopants. This result is analyzed by counting the number of Fe atoms with a certain 
magnetic moment in each configuration, then applying a Gaussian distribution around each value as a simplified representation of configuration distributions in actual 
samples. The inset of Fig. \ref{Figure2}(a) shows the resulting magnetic moment distribution on the Fe atoms for the lowest energy configurations with $x = 0.09$, 0.38 and 
0.50. For $x = 0.09$, the largest peak (generated by 22 Fe atoms) is around zero magnetic moment. Five Fe moments take values around 0.15$\mu_B$ and the others take 
values around -0.17, -0.12, 0.22, 0.30 and 0.33 $\mu_B$. 
Note that the lower concentrations have a wider distribution of peaks, whereas the higher concentration shows a clear trend towards narrowing the distribution around a 
single value.
These results clearly pose a question on the validity of any calculation performed within the virtual crystal approximation,
since through such a method only an average effect can be probed and all the local site information is lost.

We have also performed calculations adding the Hubbard $U$ term to the energy functional (DFT+U) within the simplified rotationally invariant
approach~\cite{Dudarev1998} using $U_{eff} = 2$ and 3 eV for the $2\times2\times2$--supercell.
Figure~\ref{Figure2}(b) shows the magnetic moment distribution on the Fe atoms when one Ge impurity occupies either Ga1 or Ga2 
positions. The inclusion of the Hubbard term induces a local moment on some Fe atoms that is quantitatively incorrect
in comparison to the experimental results.
This proves that the use of a semi--local functional is the correct approach to calculate the physical properties of this weakly correlated material.

\begin{figure}
\begin{center}
  \subfloat[x=0.03]{\includegraphics[width=4cm, height=4cm]{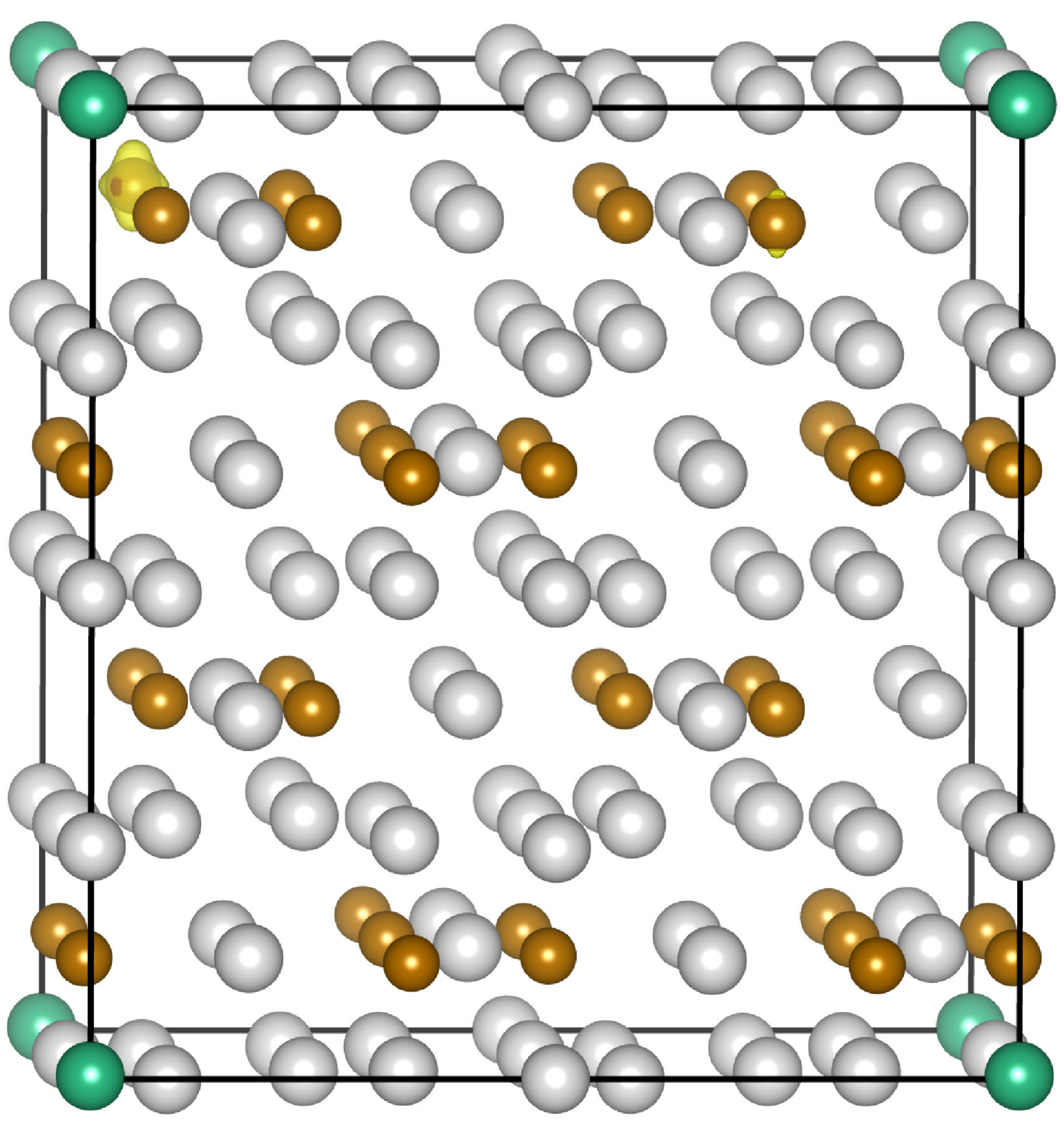}}
  \subfloat[x=0.06]{\includegraphics[width=4cm, height=4cm]{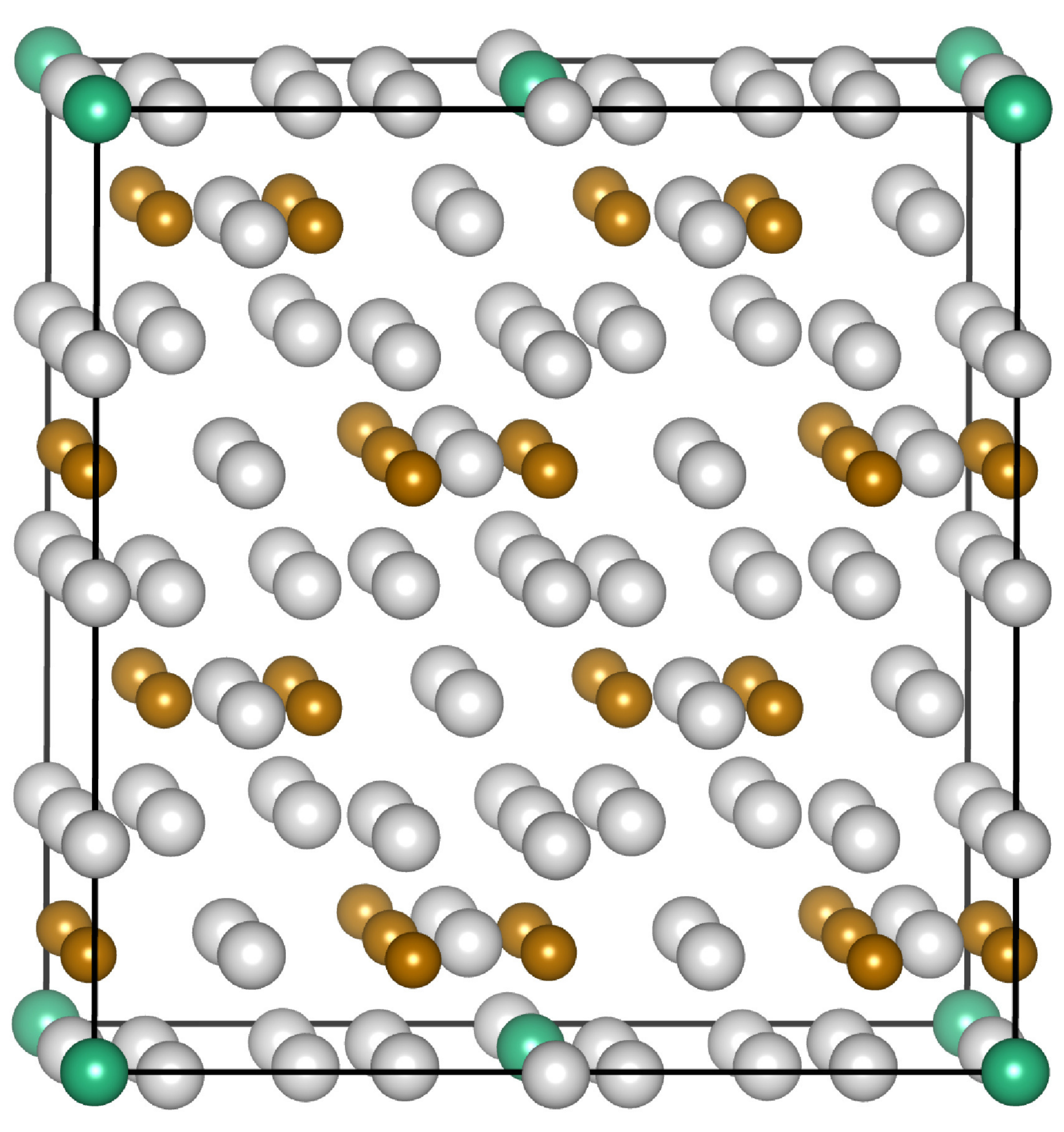}} \\
  \subfloat[x=0.09]{\includegraphics[width=4cm, height=4cm]{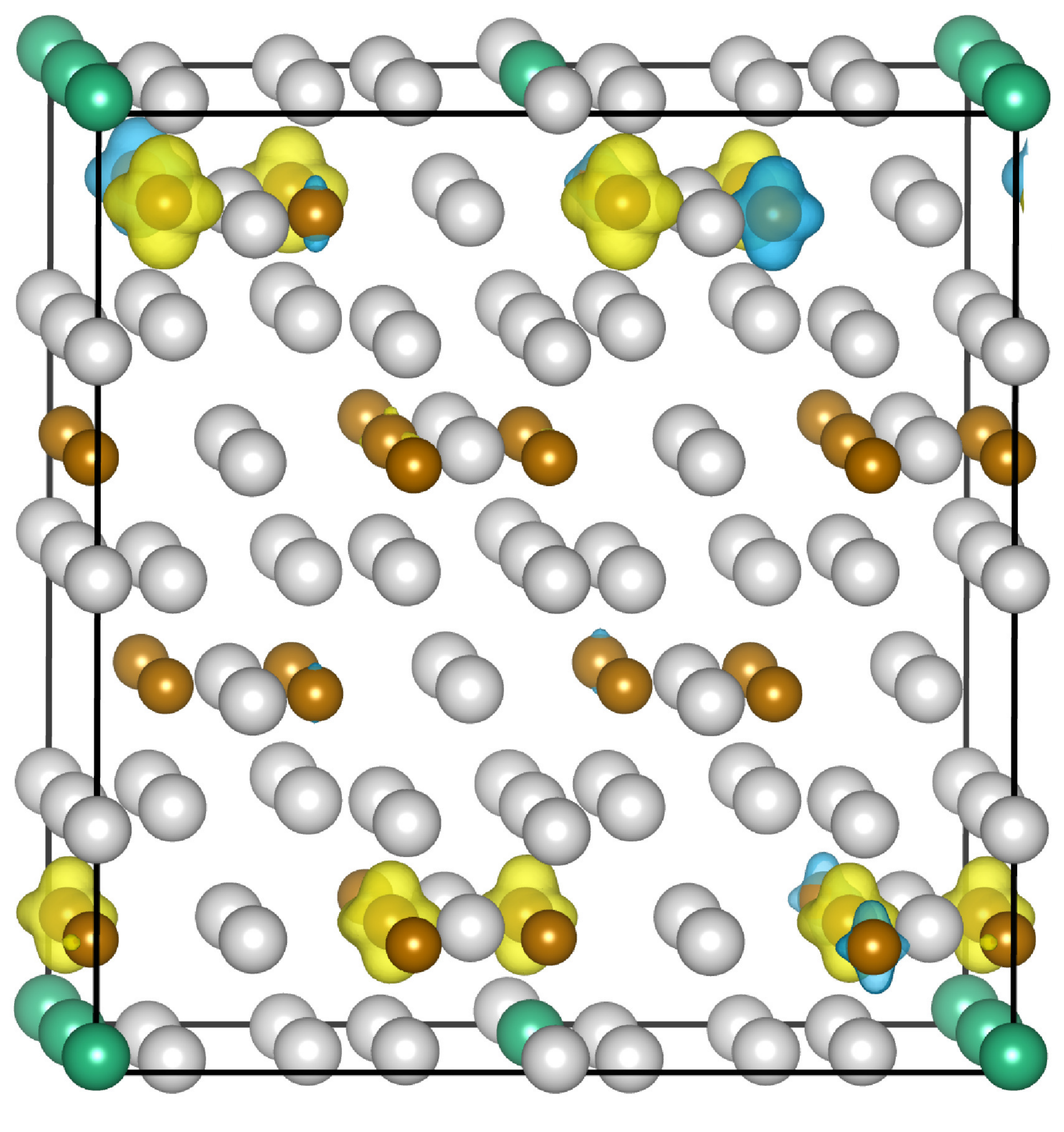}} 
  \subfloat[x=0.12]{\includegraphics[width=2cm, height=4cm]{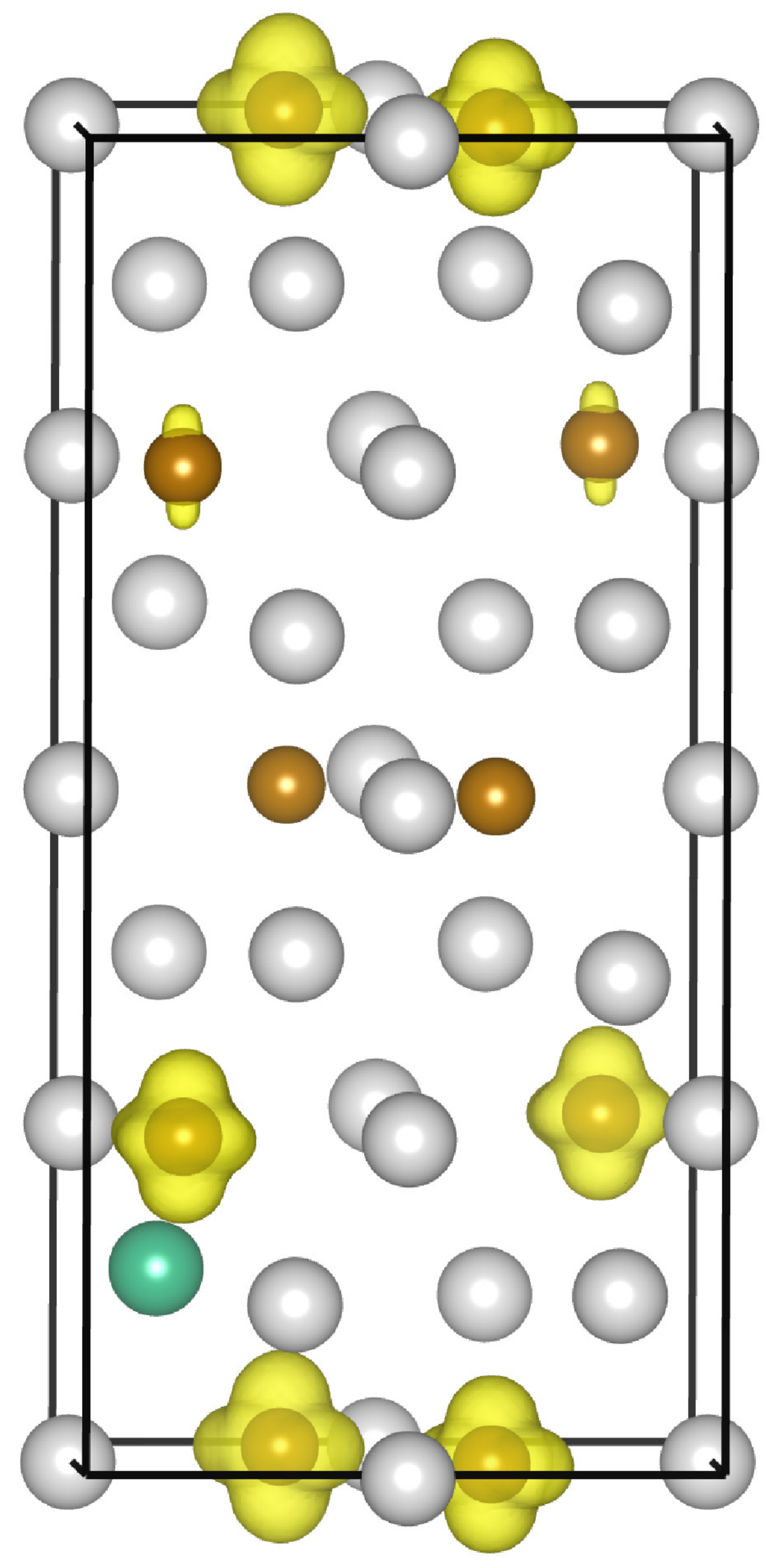}}
  \subfloat[x=0.19]{\includegraphics[width=4cm, height=4cm]{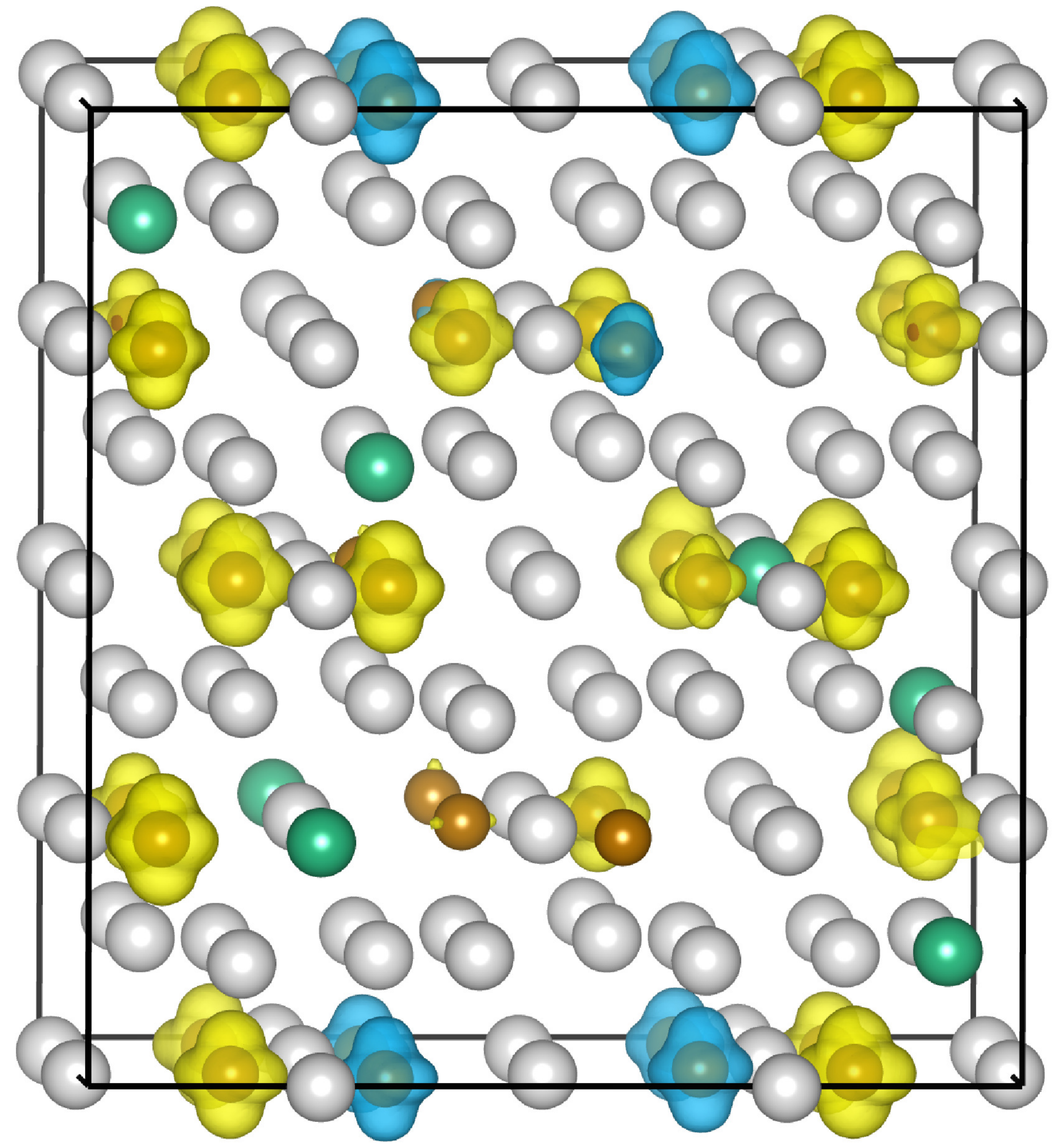}} \\
  \subfloat[x=0.25]{\includegraphics[width=2.2cm, height=2.2cm]{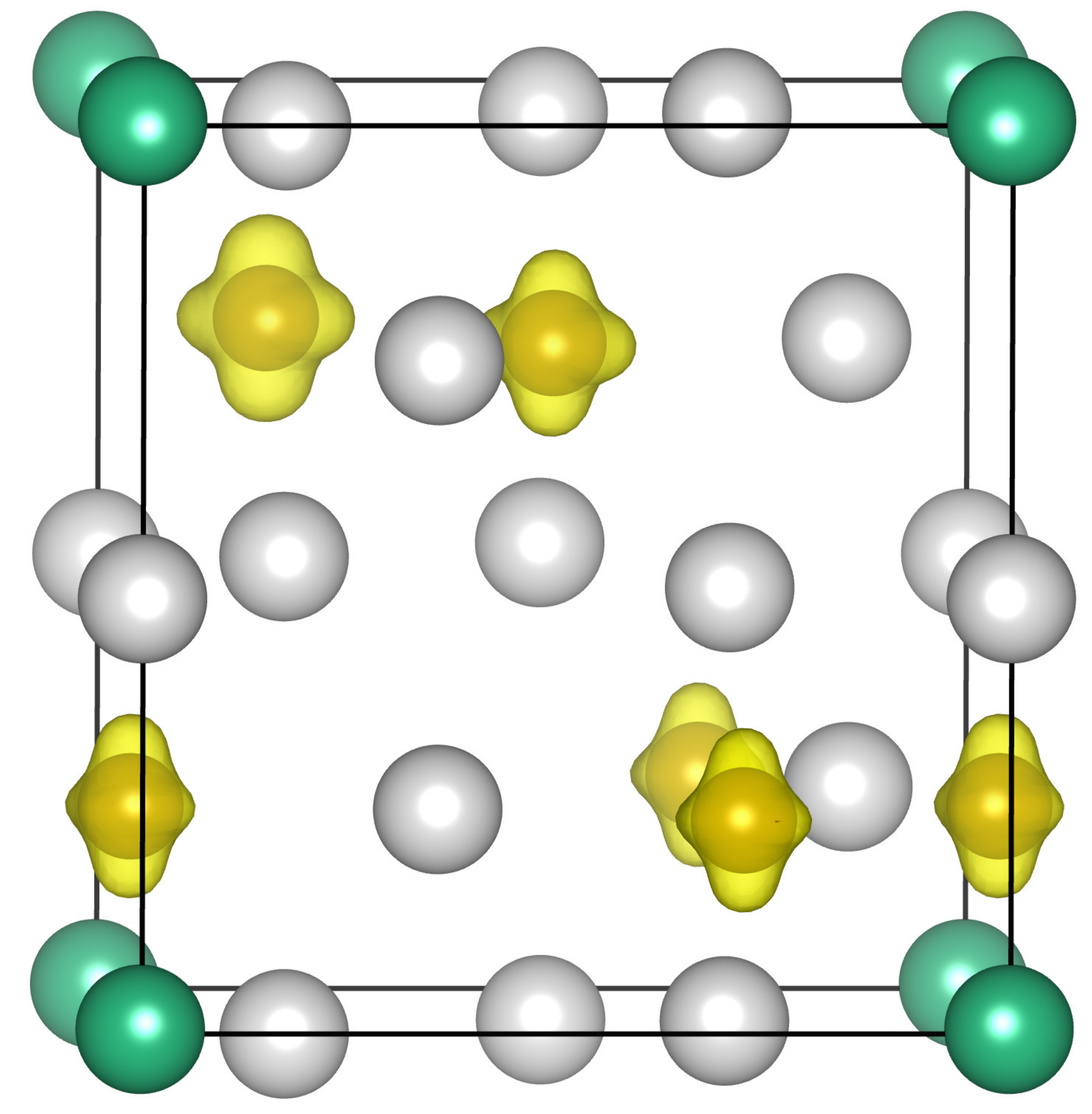} \includegraphics[width=2.2cm, height=4cm]{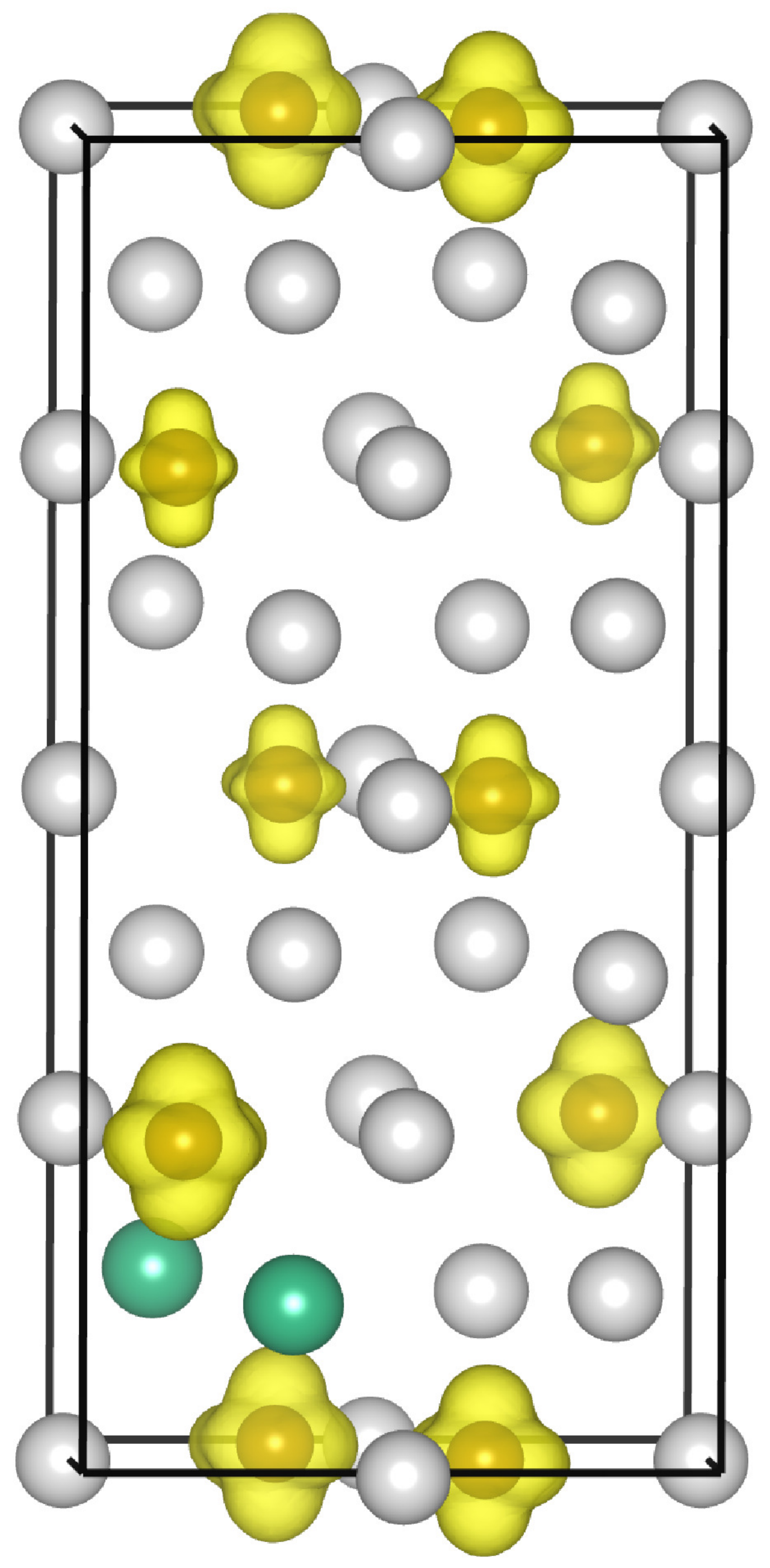} \includegraphics[width=4cm, height=4cm]{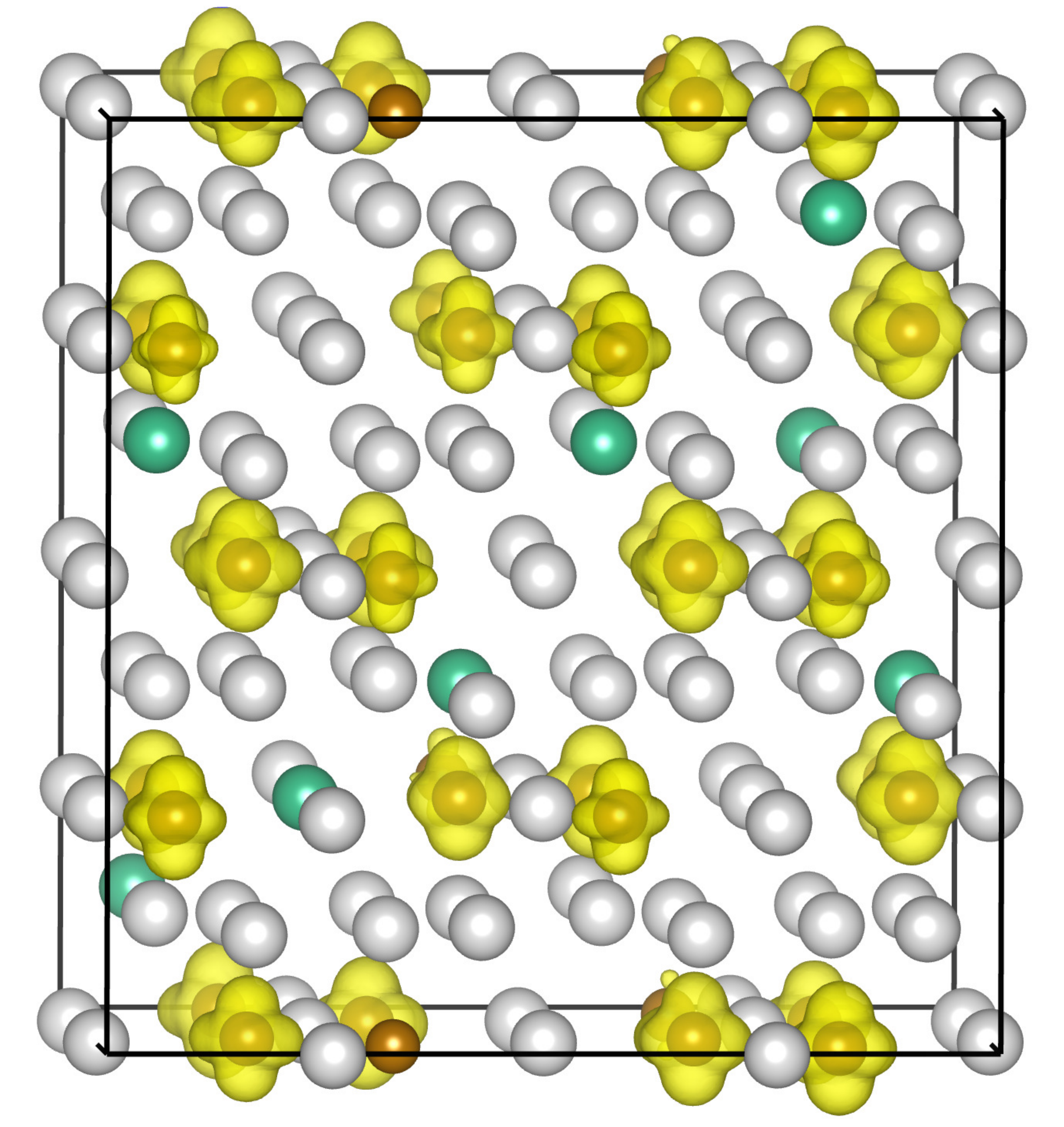}}  \\
  \subfloat[x=0.37]{\includegraphics[width=4cm, height=4cm]{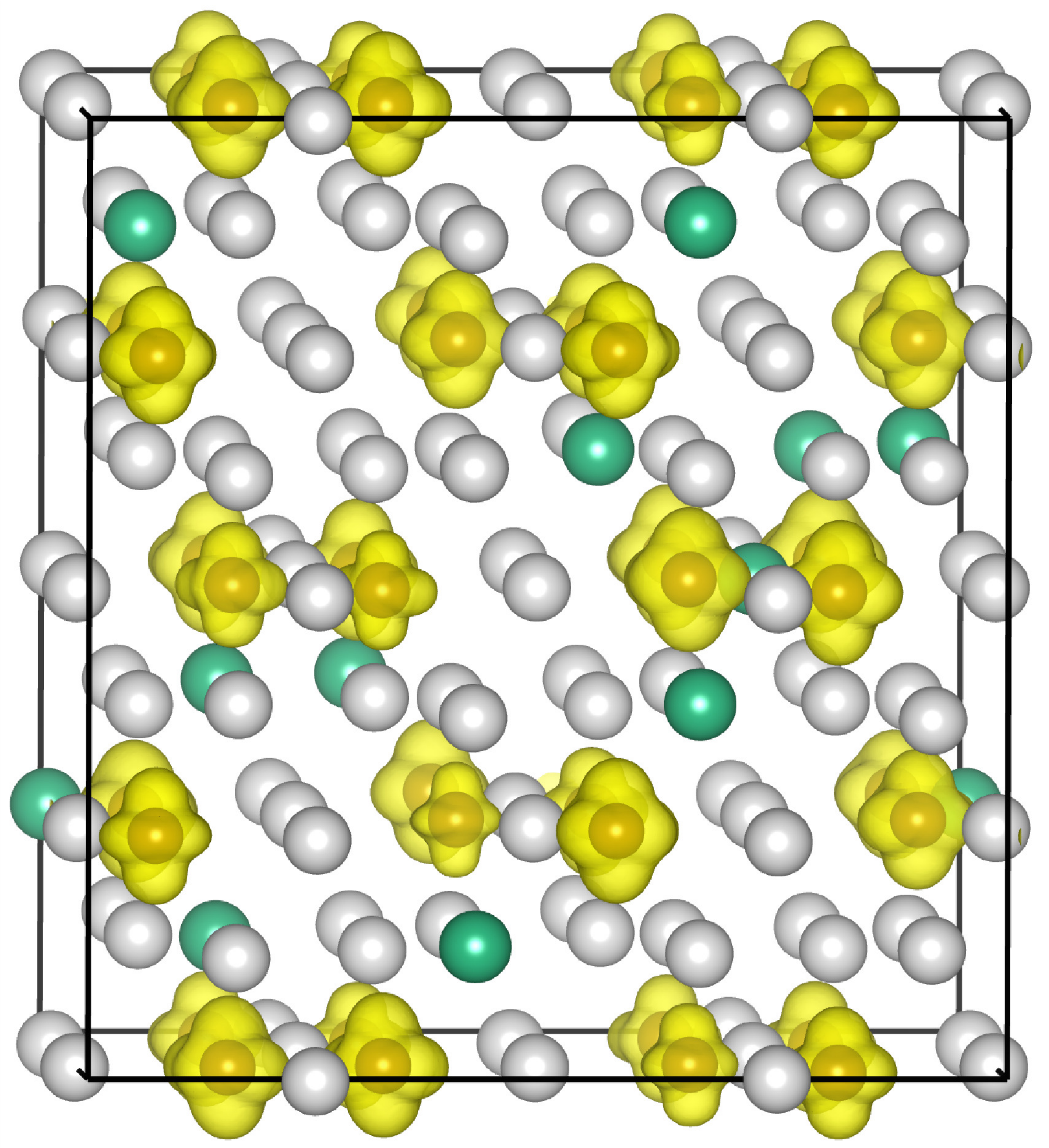}}
  \subfloat[x=0.50]{\includegraphics[width=2.2cm, height=2.2cm]{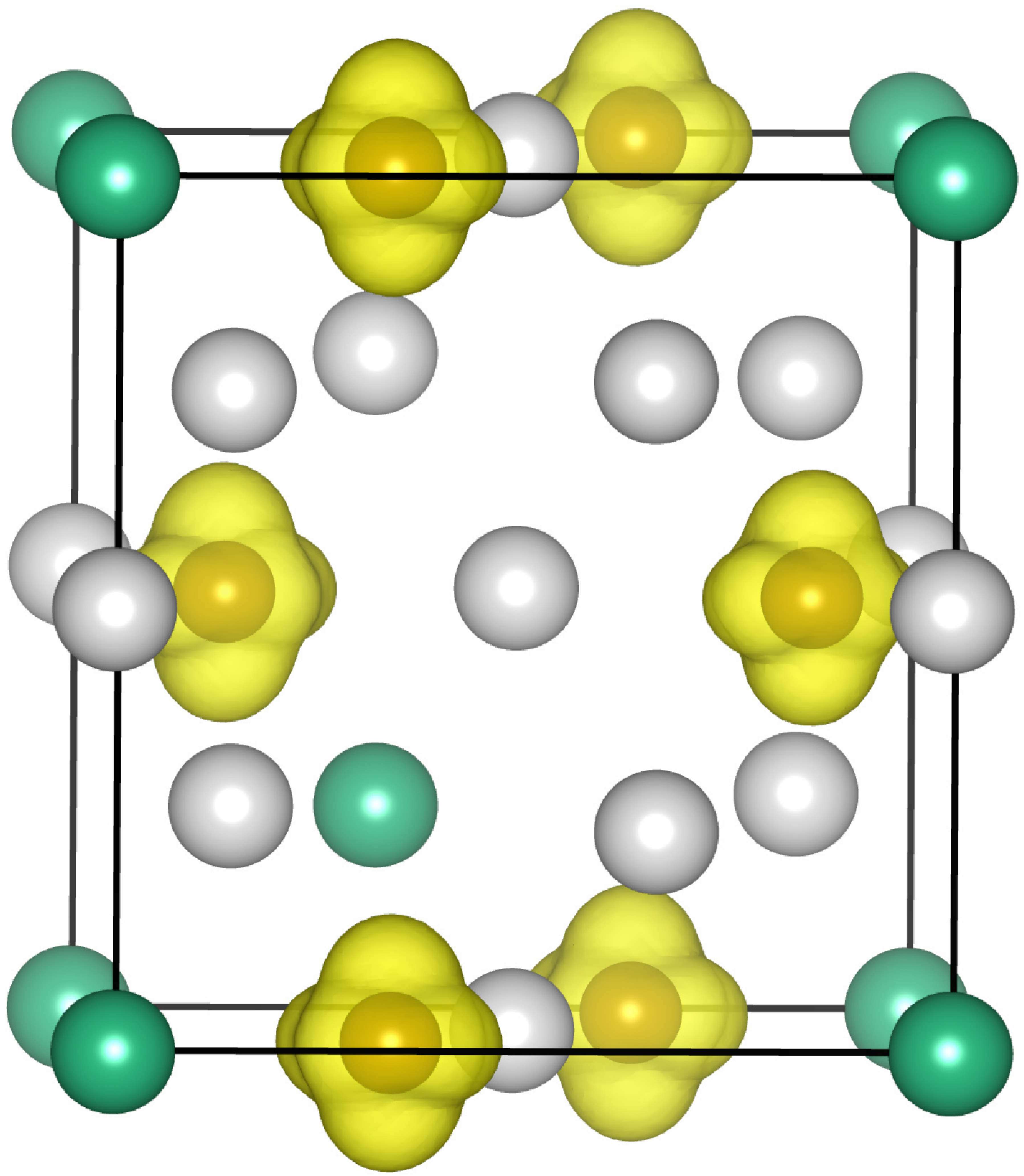}}
\end{center}
 \caption{FeGa$_{3-x}$Ge$_{x}$ magnetization density for the lowest energy configuration of some Ge concentrations $x$ and supercell sizes.
Brown, gray and green spheres represents Fe, Ga and Ge atoms, respectively.
The yellow and blue surfaces represent positive and negative values of the magnetization density.}
\label{Figure3}					
\end{figure}

With the aim of visualizing the real-space distributions of magnetic moments in the system, we plot the calculated spin densities
for the more stable distribution of Ge atoms at each concentration (Fig. \ref{Figure3}(a)-(h)). 
In these representations one can observe that the induced magnetic moments (yellow surfaces representing spin densities in an arbitrary $z$-direction) are initially localized close to the impurity as we can see in Figs. \ref{Figure3}(a)-(d). By increasing the dopant concentration $x$, the 
magnetization gradually spreads throughout the Fe atoms on the lattice, leading to a more uniform and delocalized character of
the magnetization at high concentrations as shown in Figs. \ref{Figure3}(f)-(h).
Also, our results show that the induced magnetization does not depend on the supercell size, this is clearly observe in Fig.~\ref{Figure3}(f)
($x$ = 0.25) where the magnetization densities for the unit cell, $1\times1\times2$ and $2\times2\times2$ supercells are fairly similar.
It is also worthy of notice that some of the lowest energy configurations exhibit the formation of a few antiparallel spins on the Fe dimers, exemplified in
Fig.~\ref{Figure3}(c) and (e) by the blue surfaces seen for $x=0.09$ and 0.19.
In most of the distributions
the induced magnetic moments are oriented in the same direction, leading to fully ferromagnetic states.
Thus, two main points can be derived from the PBEsol calculations which led to the results in Fig~\ref{Figure3}:
(i) an initial emergence of localized moments on some Fe atoms nearby diluted Ge impurities evolves, as a function of dopant concentration,
into a more delocalized character at larger concentrations, ending in similar moments on all Fe atoms;
(ii) in certain configurations it becomes energetically favorable for a few Fe atoms to align antiparallel to their neighbors, indicating
the presence of a minor AFM component over the majority FM interaction.

\begin{figure}
\begin{center}
\includegraphics[width=0.7\textwidth]{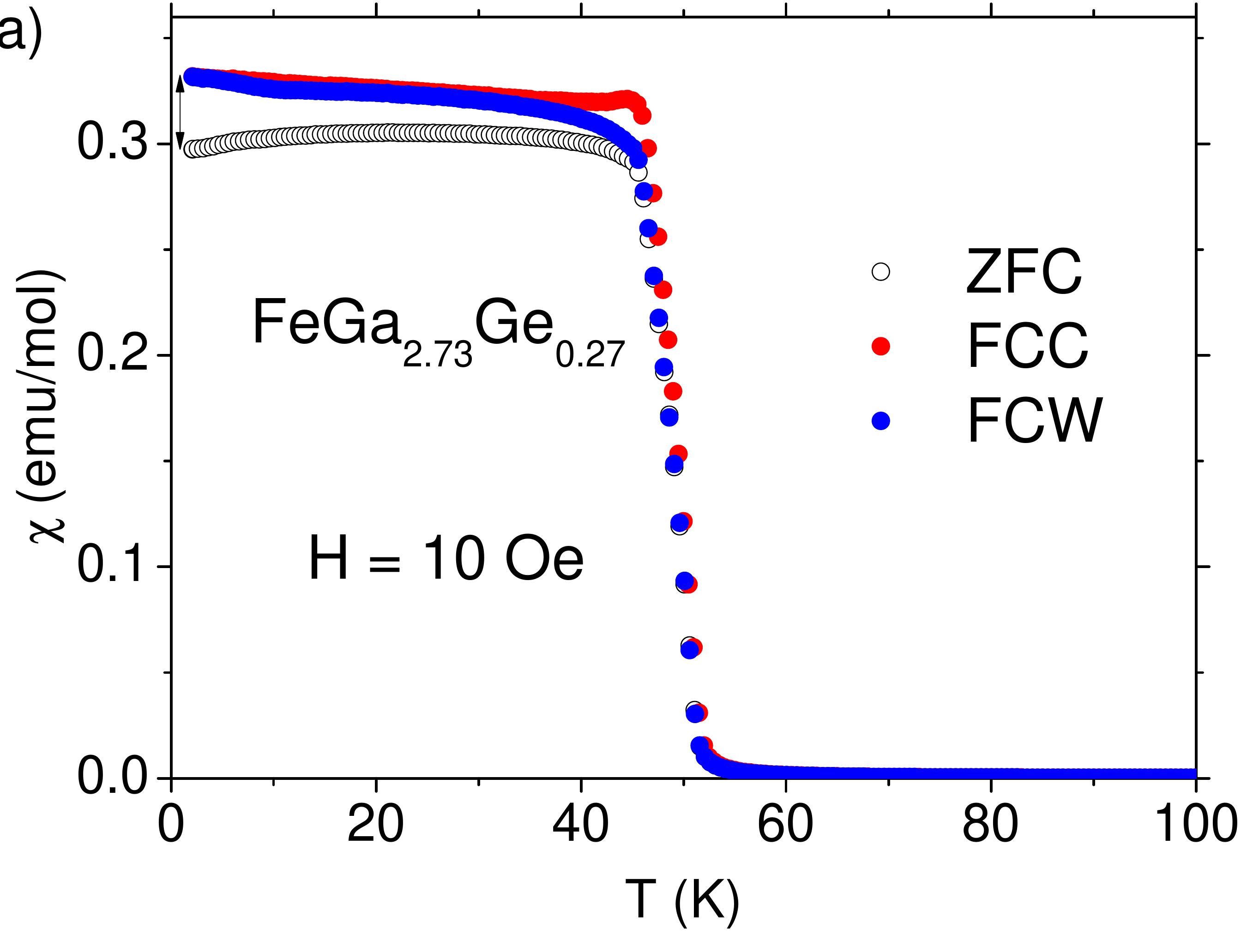}
\includegraphics[width=0.7\textwidth]{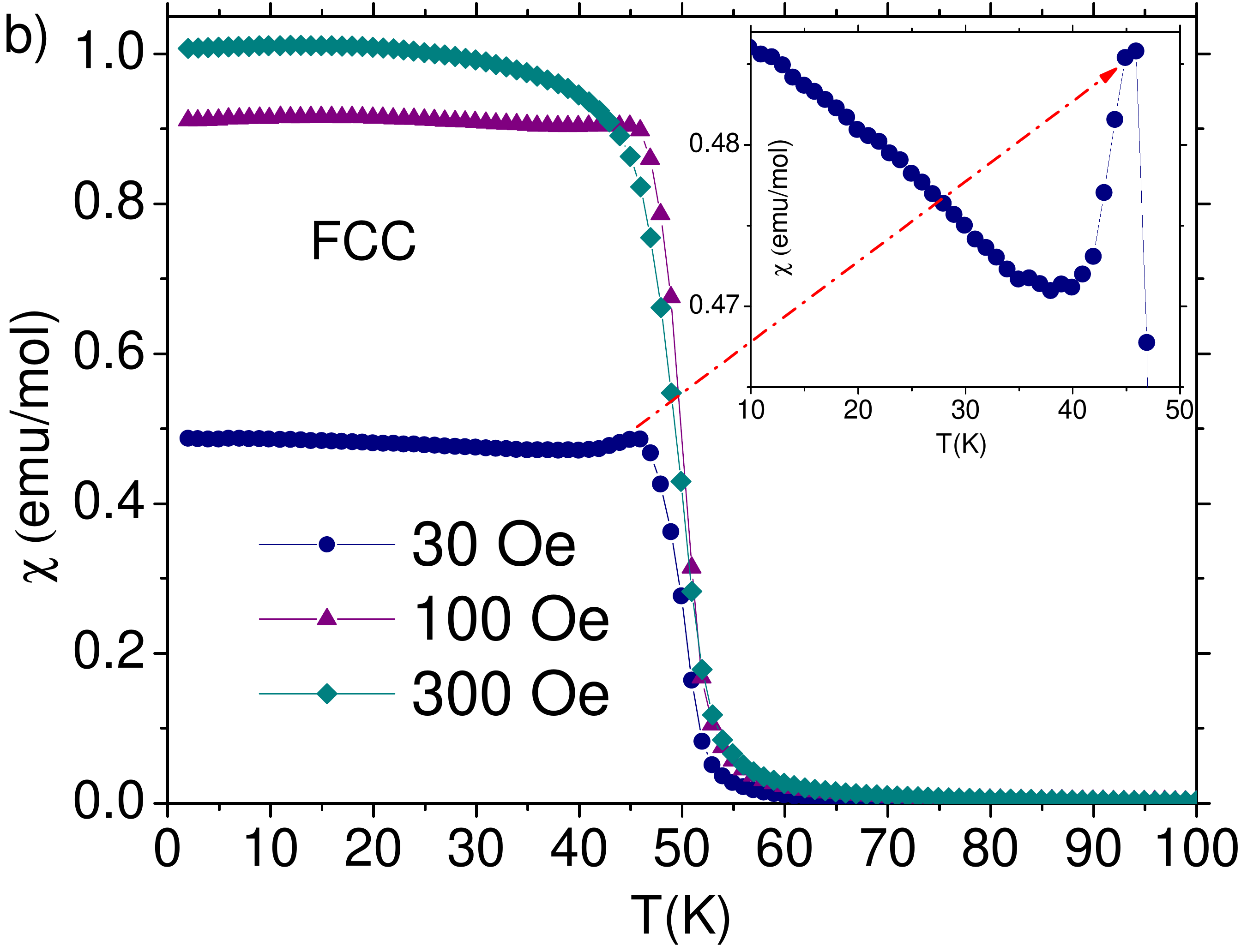}
\end{center}
\caption{a) $T$-dependence of the magnetic susceptibility at $H = 10$~Oe in the ZFC, FCC and FCW modes for FeGa$_{2.73}$Ge$_{0.27}$. b) $T$-dependence of the magnetic susceptibility at $H = 30$~Oe, 100~Oe and 300~Oe. The inset zooms in around the peak region that evidences an anomalous antiferromagnetic behavior.}
\label{Figure4}
\end{figure}

For experimental support of the above simulation results, we performed magnetic susceptibility measurements on a Ge-doped single crystal with $x=0.27$, which places it far enough into the magnetically ordered ground state region to avoid issues related to non-Fermi-liquid behavior\cite{Umeo2012}.
Fig. \ref{Figure4}(a) presents the $T$-dependence of the DC magnetic susceptibility of the FeGa$_{2.73}$Ge$_{0.27}$ crystal under external magnetic field of $H = 10$~Oe for the ZFC (black), FCC (red), and FCW (blue) branches.
In the ZFC branch the response in the saturated region (0.30 $emu/mol$) is lower than the other branches, and shows a small initial increase up to about 20~K. This may be the result of the formation of magnetic domains in the single crystal upon cooling at zero field, which are not realigned at 2~K due to the application of such a weak magnetic field. Increasing temperatures then allow some of the domains to realign themselves.
In the subsequent FCC branch, there is a small but well defined, anomalous peak in the magnetic susceptibility around 48~K in which the susceptibility increases sharply upon cooling through $T_C$, then decreases equally sharply a little but returns to a slowly increasing behavior upon further cooling below 40~K, eventually reaching 0.33 $emu/mol$ 2~K.
In this case both the magnetic domains formed at $T_C$ and the magnetic moments were already under an applied field which defined a clear symmetry breaking direction for parallel and anti-parallel alignment.
Very similar behavior is visible in previous works reporting only FCC branches \cite{Umeo2012} but was not discussed. 
This rare behavior disappears in the subsequent FCW branch, that gradually recovers similar behavior to the ZFC branch as the sample is warmed through $T_C$. 
Surprisingly, a similar behavior appears in a weak ferromagnet compound MnSi classified as a helimagnet \cite{Lamago}. This interesting feature is expected for non-centrosymmetric structures, which is not our case.

\begin{figure}
\begin{center}
\includegraphics[width=0.7\textwidth]{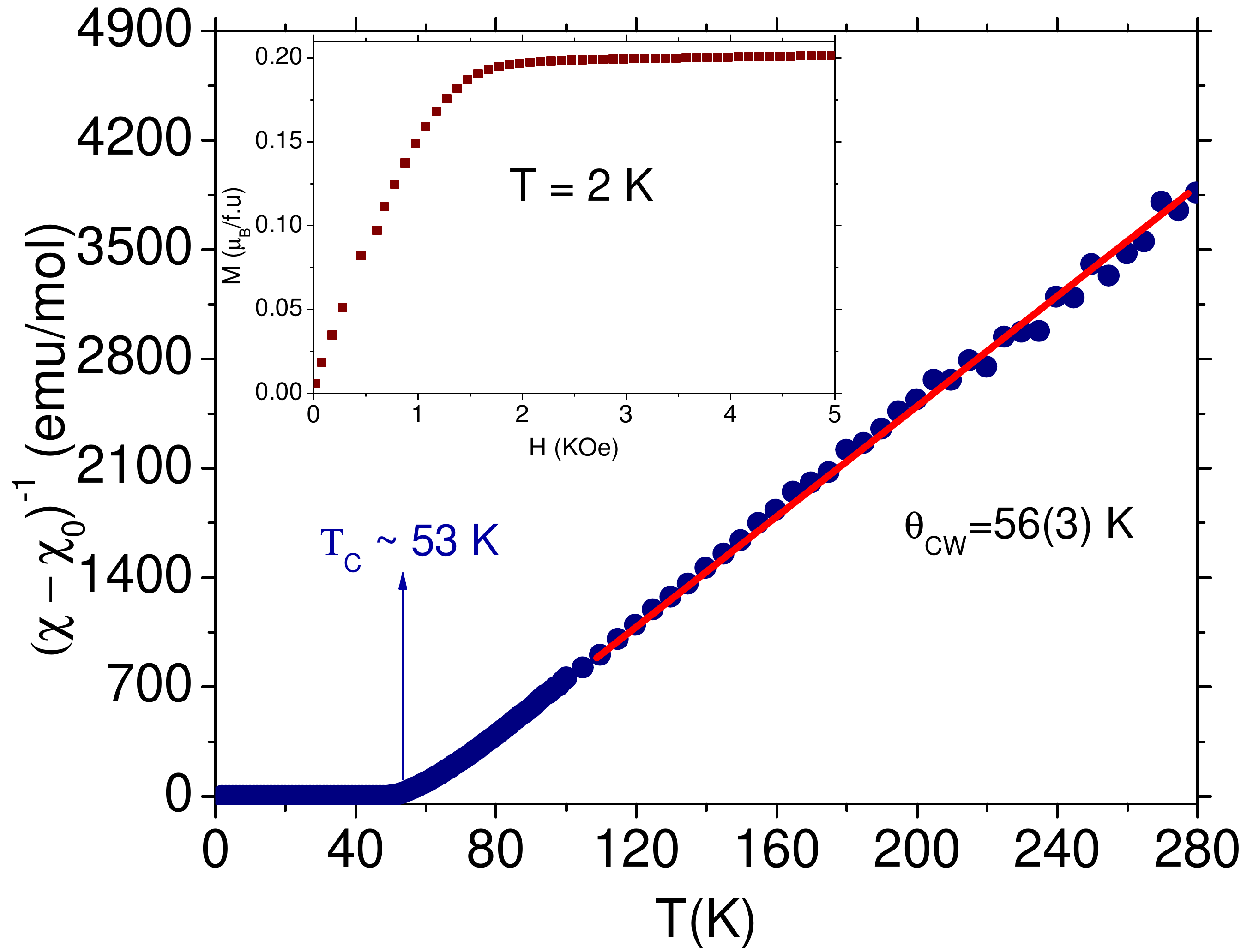}
\end{center}
\caption{Inverse susceptibility versus temperature for FeGa$_{2.73}$Ge$_{0.27}$. The red line is the fitting in high temperatures of the Curie-Weiss law. The inset shows the magnetization (M) as a function of magnetic field (H) in $T$ = 2 K.}
\label{Figure5}
\end{figure}

In order to further explore the behavior observed in the FCC curve, measurements under gradually higher applied magnetic fields ($H = 30$~Oe, 100~Oe, 300~Oe, FCC mode) were carried out and are shown in Fig. \ref{Figure4}(b).
The peak is still clearly visible at $H = 30$~Oe and detailed in the inset of Fig. \ref{Figure4}(b). Under 100~Oe the FCC peak has practically disappeared, and under 300~Oe the sample recovers the expected shape of FCC curves for traditional ferromagnets, i.e., a sharp increase around $T_C$ followed by further, gradual increase towards saturation as thermal energy is removed from the spin system.
The small but sharp decrease just below $T_C$ observed in low-field FCC curves is therefore interpreted as evidence of a minority of antiferromagnetic interactions present in the sample, resulting in some of the Fe moments aligning antiparallel to the majority, as was found in the DFT calculations for $x=0.09$ and 0.19 lowest energy configurations (Figs. \ref{Figure3}(c) and \ref{Figure3}(e)). 
Although we cannot rule out the influence of magnetic domains, we argue that it is only possible to see this weak antiferromagnetic contribution when all of the domains and spins organize under a well-defined alignment axis (FCC case).
Also, this is clearly not a trivial case of ferrimagnetism, since the antiparallel moments align themselves at a discernibly lower temperature than the majority (possibly indicating a certain level of frustration) and because the antiferromagnetic component proves to be quite fragile.
For higher external fields the peak quickly disappears and the susceptibility curves adopt the more traditional, monotonic increase upon cooling that is expected of a ferromagnet.
It is worth noting that the different values of susceptibility saturation at low temperature for these three applied magnetic fields are still compatible with an almost linear behavior of the magnetization curve at low fields, within the resolution of our experiments.
By fitting the inverse susceptibility with a modified Curie-Weiss law at high temperatures (see Fig. \ref{Figure5}), we extract $\mu_{eff} = 0.84(6) \mu_{B}$/f.u, $\theta_{CW}$ = 56(3) K and $\chi_{0}$ = 3.7(8)$\times 10^{-4}$emu/mol. 

The inset of Fig. \ref{Figure5} presents a magnetization measurement showing an almost linear behavior at low fields and no significant hysteresis within the precision of the SQUID measurement.
In the fully ferromagnetic state, the saturated magnetic moment $\mu_{sat} = 0.19(1) \mu_{B}$/f.u. at $T = 2$~K is comparable to the theoretically calculated average of 0.22 $\mu_{B}$/Fe for the more stable configuration with $x=0.25$, (see Fig. \ref{Figure2}(a)).
It is worth noting that the effective moment, $\mu_{eff}$, is different compared with the saturation moment, $\mu_{sat}$.
The itinerant vs. localized character of ferromagnets can be characterized by the Rhodes-Wolhfarth Ratio, RWR = $\mu_{eff}$/$\mu_{sat}$ \cite{Zhang2016}. For a localized system, the value of RWR should be close to 1, while it diverges for itinerant ferromagnets. 
In our case, RWR is about 4.4, supporting the scenario of intermediate behavior between localized and itinerant forms of the ferromagnetism as our DFT results showed.
Finally, the evolution from localized towards more delocalized magnetic character as a function of dopant concentration, and also the coexistence of FM and AFM components, were also supported by recent Nuclear Quadrupolar Resonance measurements \cite{Majumder2015}.

\section{Conclusions} 

In conclusion, we have shown that the magnetism of Ge-doped FeGa$_3$ is in fact much richer than what had been described so far.
Magnetic moment distributions throughout the Fe atoms are complex and strongly dependent on Ge dopant concentration, as well as on the manner in which they are 
distributed in the lattice.
For small dopant concentrations, Fe site symmetry is broken and different types of Fe atoms emerge: some bearing localized moment (induced by nearby Ge) and the rest 
remaining non spin polarized.
This scenario gradually evolves with increasing Ge concentration to a more delocalized character until the dominance of itinerant magnetism is established at the highest 
experimentally achieved doping levels ($x\sim0.41$).
Additionally, the magnetic ordering is found to be mainly FM, but in some cases it is found a minor AFM component brought in by Fe moments that align antiparallel 
under circumstances not yet fully understood (once again strongly dependent on the dopant concentration and configuration). These predictions made by our meticulous DFT 
calculations were very well supported by magnetic susceptibility measurements on a FeGa$_{2.73}$Ge$_{0.27}$ crystal, that showed a fragile AFM component 
distinguishable from the major FM transition.

What makes this fascinating system even more compelling is that all these rich features arise even before mixing in the broad range of anomalies that accompanies its reported quantum critical behavior around $x=0.13$, which is not contemplated in either the theoretical modeling or the experiments, intentionally performed here on a sample that is well into the magnetically ordered region.

\emph{Note added to arXiv version}: this paper was originally submitted as a collaboration which included preliminary $^{57}$Fe M\"ossbauer data, but after further consideration the authors involved in those experiments have chosen to prepare a separate, comprehensive publication to be submitted briefly.

\section*{Acknowledgments}
This work was supported by Brazilian agencies CAPES, CNPq and FAPESP (Contract \#'s 2011/19924-2, 2012/17569-2).
JMOG would like to thank CIEN-CODI-Vicerrector\'ia de Investigaci\'on-Universidad de Antioquia (Estrategia de Sostenibilidad 2016--2017).

\newpage

\end{document}